\documentclass[conference]{IEEEtran}
\IEEEoverridecommandlockouts
\usepackage{multirow}
\usepackage{booktabs}
\usepackage{cite}
\usepackage{amsmath,amssymb,amsfonts}
\usepackage{arydshln}
\usepackage{algorithmic}
\usepackage{graphicx}
\usepackage{threeparttable}
\usepackage{textcomp}
\usepackage{xcolor}
\def\BibTeX{{\rm B\kern-.05em{\sc i\kern-.025em b}\kern-.08em
    T\kern-.1667em\lower.7ex\hbox{E}\kern-.125emX}}
\begin{document}

\title{Revolutionizing Disease Diagnosis with simultaneous functional PET/MR and Deeply Integrated Brain Metabolic, Hemodynamic, and Perfusion Networks\\

\author{\IEEEauthorblockN{Luoyu Wang\IEEEauthorrefmark{1}\thanks{Luoyu Wang and Yitian Tao contributed equally to this work.}, Yitian Tao\IEEEauthorrefmark{1}, Qing Yang\IEEEauthorrefmark{1}, Yan Liang\IEEEauthorrefmark{1}, Siwei Liu\IEEEauthorrefmark{2}, Hongcheng Shi\IEEEauthorrefmark{2}, Dinggang Shen\IEEEauthorrefmark{1} and Han Zhang\IEEEauthorrefmark{1}\IEEEauthorrefmark{3}\IEEEauthorrefmark{4}\thanks{Corresponding author: Han Zhang, zhanghan2@shanghaitech.edu.cn }}
\IEEEauthorblockA{\IEEEauthorrefmark{1}School of Biomedical Engineering, ShanghaiTech University, Shanghai, China}
\IEEEauthorblockA{\IEEEauthorrefmark{2} Department of Nuclear Medicine, Zhongshan Hospital, Shanghai, China}
\IEEEauthorblockA{\IEEEauthorrefmark{3} State Key Laboratory of Advanced Medical Materials and Devices, ShanghaiTech University, Shanghai, China}
\IEEEauthorblockA{\IEEEauthorrefmark{4} Shanghai Clinical Research and Trial Center, Shanghai, China}
}
}

\maketitle

\begin{abstract}
   % Electroencephalography (EEG), known for its non-invasiveness, ease of use, and cost-effectiveness, has been a popular method for capturing brain signals. However, current EEG-to-Text decoding approaches face challenges due to the huge domain gap between EEG recordings and raw texts, inherent data bias, and small closed vocabularies. In this paper, we propose SEE: Semantically Aligned EEG-to-Text Translation, a novel method aimed at improving EEG-to-Text decoding by seamlessly integrating two modules into a pre-trained BART language model: (1) a Cross-Modal Codebook that learns cross-modal representations to enhance feature consolidation and mitigate domain gap, and (2) a Semantic Matching Module that fully utilizes pre-trained text representations to align multi-modal features while considering noise caused by false negatives. Experimental results on the Zurich Cognitive Language Processing Corpus (ZuCo) demonstrate the effectiveness of SEE, significantly advancing the state of EEG-to-Text decoding.
   
Simultaneous functional PET/MR (sf-PET/MR) presents a cutting-edge multimodal neuroimaging technique. It provides an unprecedented opportunity for concurrently monitoring and integrating multifaceted brain networks built by spatiotemporally covaried metabolic activity, neural activity, and cerebral blood flow (perfusion). Albeit high scientific/clinical values, short in hardware accessibility of PET/MR hinders its applications, let alone modern AI-based PET/MR fusion models. Our objective is to develop a clinically feasible AI-based disease diagnosis model trained on comprehensive sf-PET/MR data with the power of, during inferencing, allowing single modality input (e.g., PET only) as well as enforcing multimodal-based accuracy. To this end, we propose MX-ARM, a multimodal MiXture-of-experts Alignment and Reconstruction Model. It is modality detachable and exchangeable, allocating different multi-layer perceptrons dynamically ("mixture of experts") through learnable weights to learn respective representations from different modalities. Such design will not sacrifice model performance in uni-modal situation. To fully exploit the inherent complex and nonlinear relation among modalities while producing fine-grained representations for uni-modal inference, a modal alignment module is utilized to line up a dominant modality (e.g., PET) with representations of auxiliary modalities (MR). We further adopt multimodal reconstruction to promote the quality of learned features. Experiments on precious multimodal sf-PET/MR data for Mild Cognitive Impairment diagnosis showcase the efficacy of MX-ARM toward clinically feasible precision medicine.
\end{abstract}


\section{Introduction}
A clinically feasible and accurate artificial intelligence (AI)-based disease diagnosis model based on contemporary neuroimaging techniques is highly desirable for precision medicine\cite{lima2022comprehensive}. Literature has witnessed significant advances of Magnetic Resonance (MR) Imaging and Positron Emission Tomography (PET)-based individualized diagnosis, not only relieving tedious human labor but also expanding knowledge on disease mechanisms\cite{liu2023vivo}. Recently, integrated PET/MR equipment has provided a unique opportunity for revealing molecular and anatomical changes with a single scan, making PET/MR study a hot clinical research focus\cite{liu2023recent}. However, current studies either used anatomical MR with PET\cite{weber2020clinical}, or treated the two modalities as separate sources in downstream tasks\cite{huang2023automatic}. The potential of PET/MR has yet to be exploited. 

Instead of using traditional modality fusion strategy that requires full-modality data during inference, we propose a multimodal MiXture-of-experts Alignment and Reconstruction Model (MX-ARM) that adopts a modality-detachable architecture to ease full-modality requirement for inference. The Mixture-of-Experts uses a fingerprint-based router to dynamically allocate modality-specific, learnable weights ("fingerprints") for a combination of various multi-layer perceptrons ("experts"). This design closes the gap of inherent data bias among different modalities and supports uni-modal inference without sacrificing model performance, since the combination of the experts is modality independent. To fully exploit the inherent dependency among different modalities for regularization in the learning process while ensuring multimodal consistency, a modal alignment module is designed to align single modal (in this work, PET) representations with those from other auxiliary modalities (functional MRI and perfusion MRI). These designs can lead to fine-grained representations for uni-modal disease identification in the testing (clinical implementation) phase. Additionally, we adopt a multimodal reconstruction module to measure and also promote the quality of the learned representations. We test the model on a carefully curated simultaneous functional PET/MR (sf-PET/MR) dataset for early AD diagnosis.

The contributions of this work include \textit{1)} a pioneering design for multimodal brain connectome modeling in a disease population with simultaneously acquired functional PET/MR. To our knowledge, this represents the first medical image analysis study on brain metabolism, hemodynamics and perfusion; \textit{2)} A novel AI framework trained on multimodal sf-PET/MR but implemented on single modality data with carefully balanced accuracy and clinical flexibility; \textit{3)} A fingerprint-based mixture-of-experts adapter for adaptive multimodal learning and uni-modal inferencing; and \textit{4)} Modules for modality Alignment and Reconstruction to improve representation quality and promote diagnostic accuracy.

\section{Method}
\begin{figure*}[]
\centering{\includegraphics[width=0.8\textwidth]{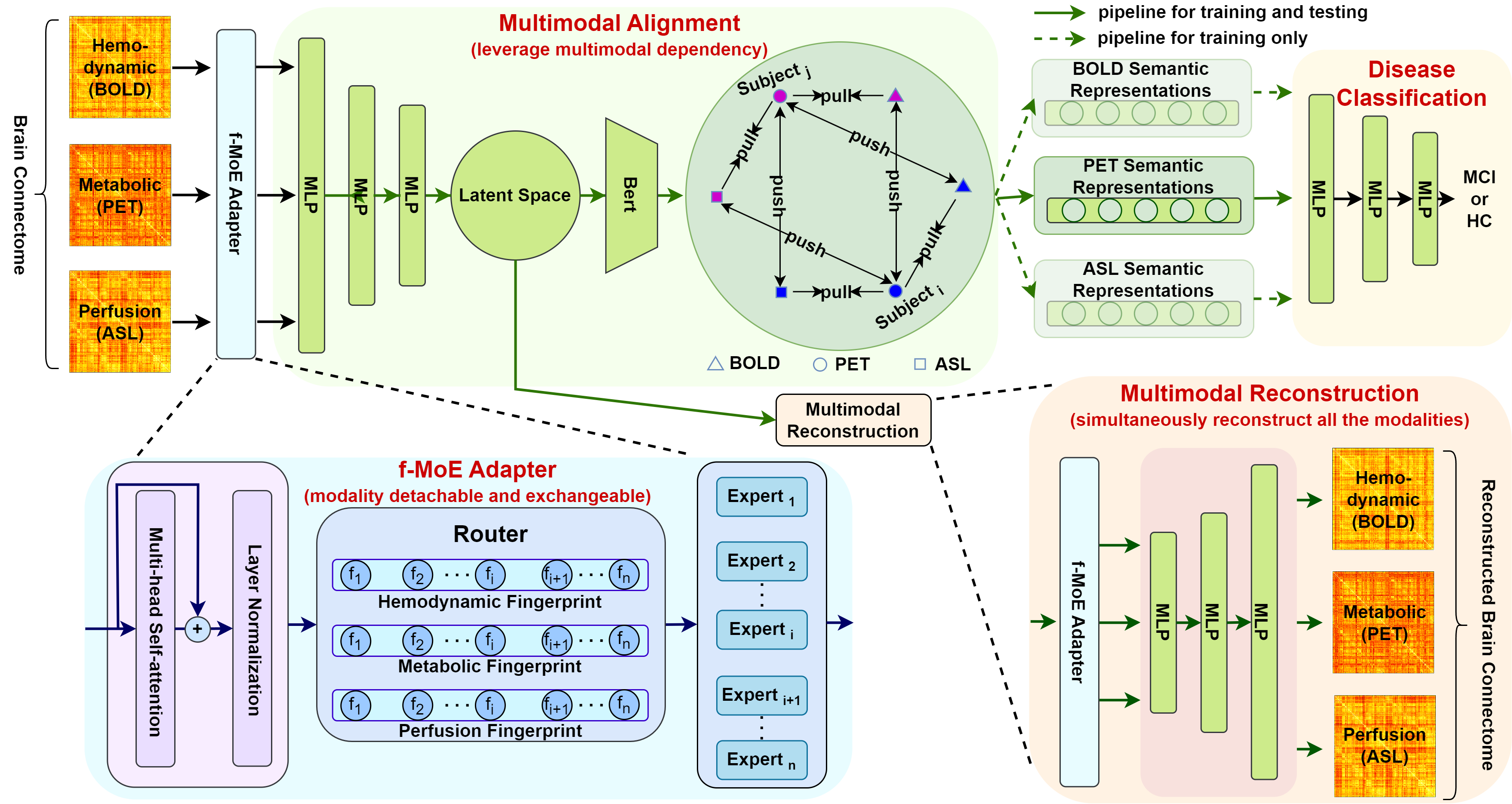}}
\vspace{-0.3cm}
\caption{Flowchart of the MX-ARM. The f-MoE Adapter uses a Router to allocate learnable modality-specific fingerprints to experts, which results in dynamic learning of multimodal representations. The Multimodal Alignment together with the Multimodal Reconstruction Module helps figure out the inter and intra-dependency of learned representations, facilitating learning fine-grained representations for disease diagnoses.} 
\label{fig1}
\vspace{-0.3cm}
\end{figure*}

% After constructing connectivity matrices, we propose our MX-ARM model aimed at MCI diagnoses. 
As shown in Fig.\ref{fig1}, MX-ARM consists of four parts: a fingerprint-based Mixture-of-Experts (f-MoE) Adapter, a MultiModal Alignment (MMA) module, a MultiModal Reconstruction (MMR) module, and a Disease Classifier. During the training phase, the input of MX-ARM is the combination of brain connectome derived from sf-PET/MR, representing spatiotemporal covariation of neural activity (i.e., blood-oxygen-level-dependence [BOLD]), metabolic activity (PET), and cerebral blood flow (i.e., perfusion, from Arterial Spin Labeling [ASL] MRI). They take the form of: $M^{BOLD}\in \mathbb{R}^{N_{b} \times N_M \times N_M}, M^{PET}\in \mathbb{R}^{N_{b} \times N_M \times N_M}, M^{ASL} \in \mathbb{R}^{N_{b} \times N_M \times N_M}$, where $N_{b}$ is the batch size and $N_M$ is the dimension of brain connectome (number of brain regions). For inference, the input is $M^{PET}$ only, since the current method integrates three modalities with PET as the main modality. However, without losing generality, it can take more modalities and treat any modality as the main modality. 
% During the inference phase, the input is PET connectivity matrices. Given the differing value ranges of the multimodal connectivity matrices, we initially employ an f-MoE Adapter for self-adaptive projections across multimodal data, creating modality-specific representations. Subsequently, these representations are projected into a latent space using a linear encoder and a Bert encoder with classification ($[CLS]$) tokens\cite{devlin2018bert}, generating semantic embeddings for each modality. These embeddings are then aligned using a triplet alignment loss, with PET-derived semantic embeddings specifically utilized for MCI classification. Furthermore, we implement a multimodal reconstruction task, compelling the model to reconstruct all modalities from each singular modality-specific representation in the latent space, which further regularizes the learning process.
Considering the heterogeneity and scale difference across modalities, we initially employ an f-MoE Adapter for self-adaptive projections across modalities. This nuanced adaptation ensures optimal interpretation  of diverse data characteristics inherent to all modalities used for training while creating modality-specific representations. Subsequently, these representations are projected into a latent space using a linear encoder and a Bert encoder with classification ($[CLS]$) tokens\cite{devlin2018bert}, generating semantic embeddings for each modality. These embeddings are then aligned using a triplet alignment loss, with main modality (PET)-derived semantic embeddings specifically utilized for disease classification. Furthermore, we incorporate a multimodal reconstruction task, compelling the reconstruction of all modalities from each modality-specific representations, which further regularizes the learning process.

\subsection{Fingerprint-based Mixture-of-Experts Adapter}
% Due to different value ranges of multimodal connectivity matrices, instead of directly fusing all the modalities together (which requires pair-wised data during the inference phase), we choose MoE module as a more flexible and also modality-detachable way to dynamically learn the deep correlation between different modalities. Different from some recent works which utilize MoE in a modality-specific or sparse routing way\cite{du2022glam,mustafa2022multimodal,riquelme2021scaling,zhou2022mixture}, we develop a fingerprint-based algorithm to dynamically allocate experts for all the modalities, which avoids using additional linear layers to calculate routing weights for experts and maintains flexibility. Specifically, in each layer of the f-MoE Adapter, for representations of each modality $M^m \in \mathbb{R}^{N_{b} \times N_M \times N_M}$ (where $m \in \{ BOLD, PET, ASL \}$ denotes the modality), we firstly leverage the self-attention mechanism followed by layer normalization\cite{vaswani2017attention} to capture the intra-correlation between the modality-specific features and generate intermediate representations:

Owing to the heterogeneity in value ranges across brain connectome derived from different modalities, instead of directly fusing all the modalities that requires full-modality data during the inference phase, we use Mixture-of-Experts (MoE) module in a more flexible and modality-detachable manner to dynamically learn deep dependency among modalities. Differed from some recent works that utilize MoE in a modality-specific or sparse routing way\cite{du2022glam,mustafa2022multimodal,riquelme2021scaling,zhou2022mixture}, we develop a \textit{fingerprint-based MoE} (\textbf{f-MoE}) algorithm to dynamically allocate experts for all the modalities, which avoids using additional linear layers to calculate routing weights for experts, thus maintaining flexibility. Specifically, in each layer of the f-MoE Adapter, for representations of each modality $M^m \in \mathbb{R}^{N_{b} \times N_M \times N_M}$ (where $m \in \{ BOLD, PET, ASL \}$ denotes modality), we first leverage self-attention followed by layer normalization\cite{vaswani2017attention} to capture intra-dependency among modality-specific features and generate intermediate representations:
\begin{equation}
    R^m = LN(M^m+Drop(Soft(\frac{M^mW^Q(M^mW^K)^T}{\sqrt{N_m}}) (M^mW^V))),
\end{equation}
where $W^Q$, $W^K$ and $W^V$ are linear transformation matrices to produce Query, Key and Value. $LN$, $Drop$ and $Soft$ are layer normalization, dropout and Softmax functions, respectively. To adaptively learn modality-specific representations, we introduce multiple multi-layer-perceptrons (MLP) to act as experts for different modalities: $Experts=\{MLP_1, MLP_2,..., MLP_{N_{MoE}}\}$. We also develop a router using modality-specific fingerprints $f^m \in \mathbb{R}^{N_{MoE}}$ to serve as learnable weights for future combination of the experts. Then, modality-specific representations $S^m$ can be obtained:
\begin{equation}
    S^m = MoE^m(R^m) = \sum_{i=1}^{N_{MoE}} f^m_i Experts_i(R^m),
\end{equation}
where $MoE^m(.)$ denotes the weighted average of the outputs of experts using $f^m$ as weights. Despite the shared experts among different modalities, the calculation of modality-specific representations is independent for each modality, which allows uni-modal calculation without degenerating model performance during the inference phase. Moreover, due to the simplicity of fingerprint-based router, MX-ARM can be naturally extended to applications with more modalities without much increase in computation cost when routing. 

\subsection{Multimodal Alignment}
After deriving modality-specific representations, it is essential to fully exploit deep entanglement between the main modality (in this paper, PET) and auxiliary modalities (e.g., BOLD and ASL) during training towards fine-grained representations for classification. Therefore, we invent a multimodal alignment module to align PET representations with those from BOLD and ASL, respectively. In practice, we use a shared linear encoder $l^{encoder}$ to encode modality-specific representations into a latent space: $P^m = l^{encoder}(S^m)$.
% \begin{equation}
%     \mathbf{S^m} = MLP(S^m).
% \end{equation}
Then, a shared Bert\cite{devlin2018bert} model is adopted to integrate the information of encoded representations and produce semantic embeddings $A^m\in\mathbb{R}^{N_{b} \times N_{latent}}$ (where $N_{latent}$ is the dimension of the latent space) using $[CLS] \in \mathbb{R}^{N_{b} \times N_{latent}}$ tokens:
\begin{equation}
    A^{m} = Bert([CLS], P^{m})_{[CLS]}.
\end{equation}
For multimodal alignment, we define the similarity scores as cosine similarity $sim^{m_A-m_B}_{ij} = cos(A^{m_A}_i, A^{m_B}_j)$(where $m_A$ and $m_B$ denotes two different modalities), and design a triplet alignment loss inspired by contrastive learning\cite{chen2020simple,girdhar2023imagebind,radford2021learning}:
% \begin{equation}
%     sim^{PET-ASL}_{ij} = cos(A^{PET}_i, A^{ASL}_j),
% \end{equation}
\begin{equation}
\vspace{-0.5cm}
\begin{split}
         L_{align}=-\frac{1}{N_{b}} \sum_{i=1}^{N_{b}}(\log (exp \ sim^{PET-BOLD}_{ii}+\\
         exp \ sim^{PET-ASL}_{ii})/ \tau
         -\log(\sum_{j=1}^{N_{b}}(exp \ sim^{PET-BOLD}_{ij}\\ 
         + exp \ sim^{PET-ASL}_{ij})/ \tau)).
\end{split}
\end{equation}

         % L_{align}=-\frac{1}{N_{b}} \sum_{i=1}^{N_{b}} \\
         % log\frac{(exp \ sim^{PET-BOLD}_{ii}+exp \ sim^{PET-ASL}_{ii})/ \tau}{\sum_{j=1}^{N_{b}}(exp \ sim^{PET-BOLD}_{ij} + exp \ sim^{PET-ASL}_{ij})/ \tau}.
% where $cos(.)$ denotes cosine similarity. By minimizing the alignment loss, multimodal semantic embeddings coming from the same subject are well aligned (pulled together) and those from different subjects are pushed apart, which contributes to the learning of representations and facilitates downstream tasks.
As such, multimodal semantic embeddings coming from the same subject will be pulled together (thus promoting modal alignment) and those from different subjects will tend to be pushed apart, which benefits learning of modality-independent, disease-related representations and facilitates downstream tasks.

\subsection{Multimodal Reconstruction}
Different from traditional reconstruction tasks which reconstruct only one modality at a time\cite{he2022masked}, to facilitate quality of the learned multimodal representations, we perform simultaneous multimodal reconstruction with an f-MoE Adapter (to avoid ambiguity, we denote the f-MoE Adapter used in this module as f-MoE Decoder Adapter) combined with a shared linear decoder $l^{decoder}$. For each modality, we adaptively reconstruct all the modalities at once using  $P^m$:
\begin{equation}
    C^{m-m'} = l^{decoder}(MoE^{m'}(Encoder(P^m))),
\end{equation}
% \begin{equation}
%     C^{m-BOLD} = l^{decoder}(MoE^{BOLD}(Encoder(P^m))),
% \end{equation}
% \begin{equation}
%      C^{m-PET} = l^{decoder}(MoE^{PET}(Encoder(P^m))), 
% \end{equation}
% \begin{equation}
%     C^{m-ASL} = l^{decoder}(MoE^{ASL}(Encoder(P^m))).
% \end{equation}
where $m' \in \{ BOLD, PET, ASL \}$. Practically, we use less MLPs for reconstruction, which indicates a smaller $N_{MoE}$ during reconstruction. For optimizing the reconstruction process, element-wise Mean-Squared Error (MSE) between the reconstructed and the original brain connectome is used:
\begin{equation}
\begin{split}
    L_{recon}^m = (M^{BOLD} - C^{m-BOLD})^2 + \\
    (M^{PET} - C^{m-PET})^2 + (M^{ASL} - C^{m-ASL})^2,
\end{split}
\end{equation}
\begin{equation}
    L_{recon} = L_{recon}^{BOLD} + L_{recon}^{PET} + L_{recon}^{ASL}.
\end{equation}

\subsection{Disease Classification}
After obtaining fine-grained representations (i.e., semantic embeddings), a linear classifier $l^{classifier}$ is adopted for classification: $\hat{y}^m = Sigmoid(l^{classifier}(A^m))$, with Binary CrossEntropy loss (BCE) as the classification loss:
\begin{equation}
    L_{cls} = BCE(y, \hat{y}^{BOLD}) + BCE(y, \hat{y}^{PET}) + BCE(y, \hat{y}^{ASL}),
\end{equation}
where $y\in \{0, 1\}$ is the label that indicates whether the subject is diagnosed as patient or not. Collectively, our model is trained by optimizing the joint loss:
\begin{equation}
    L = L_{align} + L_{cls} + L_{recon}.
\end{equation}

\section{Experiments}
\subsection{Datasets and Implementation Details}

For data augmentation, we use additional  PET and fMRI data  of 819 patients with Mild Cognitive Impairment (MCI) which may progress to AD and 518 healthy controls (HC) collected from ADNI3 dataset\cite{weiner2017alzheimer} for sequential training the base model. Then, we train MX-ARM on a precious sf-PET/MR in-house dataset consisting of simultaneously acquired brain functional (metabolic, hemodynamic, and perfusion) images from 48 MCI  and 62 matched HC. This is the unprecedented data concurrently monitoring AD-related brain function and connectivity changes, collected in a single scan, from a cutting-edge integrated PET/MR scanner. The PET tracer used is 18F-fluorodeoxyglucose. MCI diagnosis was conducted by neurologist adhering to the established criteria\cite{albert2011diagnosis}. All subjects provided written informed consent.

Each modality undergoes standard preprocessing detailed elsewhere\cite{esteban2019fmriprep,routier2021clinica,wang2008empirical}. The Schaefer’s atlas is used to parcellate the brain into 400 regions of interest, serving as nodes of brain connetome.  Brain metabolic and perfusion connectome are constructed by measuring the Kullback–Leibler divergence for each pair of brain regions between region-wise distributions of relative standard uptake value and cerebral blood flow, respectively\cite{wang2020individual}. Hemodynamic connectome is built using Pearson’s correlation between BOLD signals\cite{zhou2020toolbox}. Such multimodal connectome building transforms initial raw data into brain networks carrying disease-sensitive information. 
% The dataset is divided into a 7:3 ratio for training and testing. The efficacy of our model is assessed via metrics including Area Under Curve  (AUC), accuracy (ACC), F1-score (F1), sensitivity (SEN) and specificity (SPE). During the training phase, multimodal brain connectome is used; For testing, only PET connectome is used. We use a 3-layer f-MoE Adapter with $N_{MoE}$ set to 12 and a 1-layer f-MoE Decoder Adapter with its $N_{MoE}$ set to 3. $N^M$ is 400 and $N_{latent}$ is 128. The learning rate is set to $5\times10^{-5$ and the batch size is 36. The model size is 20M and it is trained on an A100 GPU using Adam as the optimizer.
% The dataset is divided into a 6: 1: 3 ratio for training, validation and testing. The evaluation metrics for model assessment are Area Under Curve (AUC), accuracy (ACC), F1-score (F1), sensitivity (SEN) and specificity (SPE). During training phase, multimodal brain connectome is used; For testing, only PET connectome is used. We use a 3-layer f-MoE Adapter with $N_{MoE}$ set to 12 and a 1-layer f-MoE Decoder Adapter with its $N_{MoE}$ set to 3. $N^M$ is 400 and $N_{latent}$ is 128. Learning rate is set to $5\times10^{-5}$ and the b

The dataset (merged by our in-house and ADNI dataset) is divided into a 6: 1: 3 ratio for training, validation and testing. The evaluation metrics for model assessment are  Area Under Curve (AUC), accuracy (ACC), F1-score (F1), sensitivity (SEN) and specificity (SPE). During training phase, multimodal brain connectome is used; For testing, only PET connectome is used. We use a 3-layer f-MoE Adapter with $N_{MoE}$ set to 12 and a 1-layer f-MoE Decoder Adapter with its $N_{MoE}$ set to 3. $N^M$ is 400 and $N_{latent}$ is 128. Learning rate is set to $5\times10^{-5}$ and the batch size is 36. The model is trained on an A100 GPU using Adam as the optimizer.

\subsection{Results and Discussion}
% \begin{table}
%   \setlength{\tabcolsep}{5pt}
%   \caption{Comparison with different classification methods. LR, SVM, GCN denotes Linear Regression, Support Vector Machine and Graph Neural Network, respectively.}
%   \label{table:1}
%   \centering
%   \begin{tabular}{ccccccc}
%     \toprule
%      Method & accuracy & AUC & sensitivity & specificity & F1  \\
%     \midrule
%     \texttt{LR} & 0.600 & 0.582 & 0.375 & 0.789 & 0.462\\
%     \texttt{SVM} & 0.629 & 0.609 & 0.375 & 0.842 & 0.480\\
%     \texttt{GNN} & 0.709 & 0.674 & 0.395 & 0.951 & 0.614\\
%     \texttt{Bert\cite{devlin2018bert}} & 0.685 & 0.666 & 0.437 & 0.894 & 0.560\\
%     \texttt{Ours} & \textbf{0.828} & \textbf{0.827} & 0.842 & 0.812 & \textbf{0.812}\\
%     \bottomrule
%   \end{tabular}
% \end{table}

\begin{table}[]
% \begin{threeparttable}
  \setlength{\tabcolsep}{5pt}
  \caption{Model performance comparison.}
  \vspace{-0.2cm}
  \label{table:1}
  \centering
  \begin{tabular}{ccccccc}
    \toprule
     Method & AUC & ACC & F1 & SEN & SPE  \\
    \midrule
    Linear Regression & 0.600 & 0.597 & 0.562 & 0.631 & 0.562\\
    Support Vector Machine\cite{khazaee2015identifying} & 0.655 & 0.657 & 0.625 & 0.625 & 0.684\\
    Graph Neural Network\cite{song2019graph} & 0.657 & 0.649 & 0.562 & 0.736 & 0.600\\
    Bert\cite{devlin2018bert} & 0.678 & 0.678 & 0.723 & \textbf{0.839} & 0.517\\
    MX-ARM (Ours) & \textbf{0.741} & \textbf{0.741} & \textbf{0.733} & 0.714 & \textbf{0.767}\\
    \bottomrule
    \end{tabular}
    \vspace{-0.2cm}
    % \begin{tablenotes} 
    % \item Note: LR: Linear Regression, SVM: Support Vector Machine, GNN: Graph Neural Network
    %  \end{tablenotes}
% \end{threeparttable}
\end{table}
% Table \ref{table:1} shows the comparison results between our method and different classification methods, our model exceeds all the other uni-modal methods with a large margin, which demonstrates the superior performance of our approach. 
% \textbf{Results. }The comparative analysis, as depicted in Table \ref{table:1}, highlights that our model outperforms traditional uni-modal methods, underlining our strategy's enhanced efficacy. An ablation study further delineates the contributions of our proposed modules in Table \ref{table:2}: the baseline being a transformer model mirroring our architecture in layer count. The abbreviations MOE, ALIGN, RECON denote f-MoE Adapter, MA and MR Modules, respectively. Insights from Trials 1 and 8 elucidate that these specialized modules collectively contribute to substantial improvements across all metrics, affirming their integral role in model refinement.
% The comparative analysis, as depicted in Table \ref{table:1}, highlights that our model outperforms traditional uni-modal methods, underlining our strategy's enhanced efficacy. An ablation study further delineates the contributions of our proposed modules in Table \ref{table:2}: the baseline being a transformer model mirroring our architecture in layer count. Insights from Trials 1 and 8 elucidate that these specialized modules collectively contribute to substantial improvements across all metrics, affirming their integral role in model refinement.
Table \ref{table:1} shows the comparison results of our method with traditional models. Note that MX-ARM uses multimodal data in training and PET data only in testing; while the competing methods use similar architectures with modality-specific parameters for representation learning and simply concatenate the modality-specific representations for disease classification in training and testing (i.e., use all modalities in testing). Our model not only outperforms other methods but also better suits for the clinical setting with uni-modal inference ability. 
Table \ref{table:2} summarizes ablation studies with the baseline being a transformer model without f-MOE Adapter, MMA and MMR modules. Insights from experiments 1 and 8 elucidate that the specialized modules collectively contribute to substantial improvement across all main metrics.
% Table \ref{table:2} illustrates the ablation study of our proposed modules, The BASE method in the table is a raw transformer model which has the same number of layers as our approach. MOE, ALIGN, RECON denote f-MoE Adapter, multimodal Alignment (MA) and multimodal Reconstruction (MR) Modules, respectively. In Trials 1 and 8, we can learn that with the help of our designed modules, a large increase in all metrics can be achieved, which illustrates the effectiveness of these modules. 
\begin{table}[]
\vspace{-0.2cm}
  \setlength{\tabcolsep}{3pt}
  \caption{Ablation studies.}
  \vspace{-0.2cm}
  \label{table:2}
  \centering
  \begin{tabular}{ccccccc}
    \toprule
    Experiment & Method & AUC & ACC & F1 & SEN & SPE   \\
    \midrule
    1 & Base & 0.666 & 0.685 & 0.560 & 0.437 & \textbf{0.894}\\  %& 0.685 & 0.666 & 0.437 & 0.894 & 0.560
    2 & Base + MMA & 0.572 & 0.600 & 0.363 & 0.250 & \textbf{0.894}\\
    3 & Base + MMA + MMR & 0.603 & 0.628 & 0.434 & 0.312 & \textbf{0.894} \\
    4 &Base + f-MoE & 0.714& 0.714   & 0.719 & 0.732 & 0.696 \\
    5 & Base + MMR & 0.659 & 0.657 & 0.647 & 0.687 & 0.631 \\ %0.657 0.659 0.631 0.687 0.666
    6 & Base + f-MoE + MMA & 0.723& 0.723   & 0.704 & 0.660 & 0.785 \\
    7 & Base + f-MoE + MMR & 0.732 & 0.732 & 0.732 & \textbf{0.732} & 0.732 \\
    % 8 & Base + f-MoE + MMA + MMR & \textbf{0.827} & \textbf{0.828} & \textbf{0.812} & 0.842 & 0.812 \\
    8 & MX-ARM & \textbf{0.741} & \textbf{0.741} & \textbf{0.733} & 0.714 & 0.767 \\
    \bottomrule
  \end{tabular}
% \begin{tablenotes} 
%     \item Note: Base, a transformer model; MMA, MMR, and f-MoE denote Multimodal Alignment, Multimodal Reconstruction, and f-MoE Adapter, respectively. MX-ARM is the combination of Base, MMA, MMR, and f-MoE modules.
% \end{tablenotes}
\vspace{-0.4cm}
\end{table}
% \vspace{-0.5cm}

\textbf{The Effect of f-MoE Adapter.}
Insights from the Experiments 1, 4, 6, 7, and 8 elucidate that \textit{1)} The implementation of the f-MoE Adapter significantly enhances model performance (Experiment 1 \textit{vs.} 4). This underscores the effectiveness of adaptive learning across different modalities; \textit{2)} With f-MoE, both MMA and MMR modules make additional contribution (Experiment 4 \textit{vs.} 6, 7), which implies that fine-grained representations are achieved by both exploiting inherent multimodal dependency and enforcing the reconstruction quality; \textit{3)} The concurrent usage of f-MoE, MMA and MMR achieves more balanced results, indicating more robust and stable learning.
% \begin{figure}[h]
% \centerline{\includegraphics[width=0.9\textwidth]{visualizarion2.png}}
% \caption{The T-SNE visualization of the PET semantic embeddings, the red points are subjects diagnosed as MCI while the blue points are HCs.} 
% \label{fig2}
% \end{figure}
% \vspace{-0.2cm}
\begin{figure}[h]
\vspace{-0.2cm}
\centerline{\includegraphics[width=0.45\textwidth]{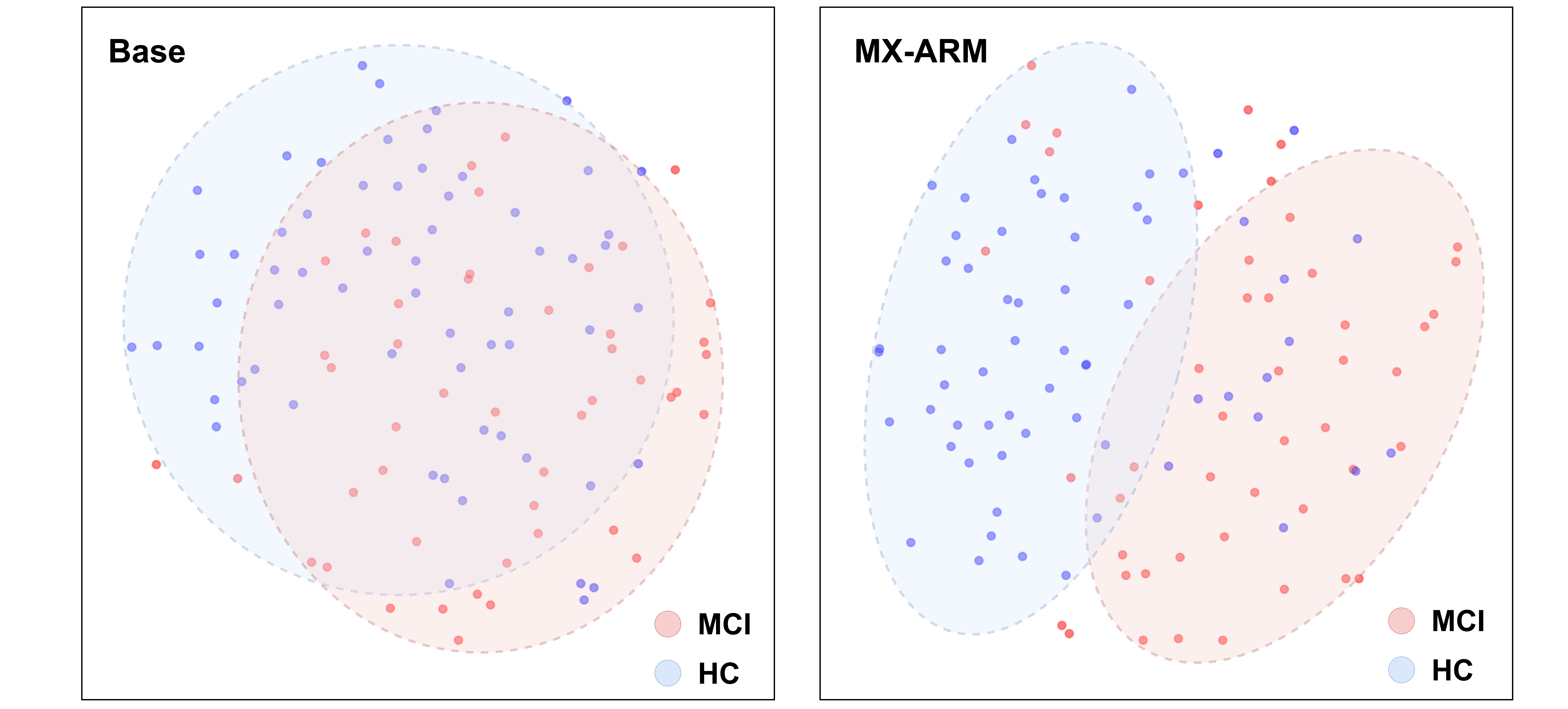}}
\vspace{-0.2cm}
\caption{The \textit{t}-SNE visualization of the PET semantic embeddings.} 
% \caption{The \textit{t}-SNE visualization of the PET semantic embeddings, the red points are subjects diagnosed as MCI while the blue points are HC.} 
\label{fig2}
\vspace{-0.2cm}
\end{figure}

\textbf{The Effect of MMA and MMR.}
From the results in Experiments 1-3, 5-7, specific roles of the MMA and MMR modules are revealed: \textit{1)} Without f-MoE, MMA seems not to work properly if used separately from MMR (Experiments 1, 2, 6). This indicates that heterogeneity of multimodal brain connectome indeed creates a bias that cannot be solved by shared linear transformation, thus results in misalignment; \textit{2)} Compared with MMA, MMR performs better if used alone (Experiments 1, 5), as the f-MoE decoder Adapter takes effect. 
% \begin{table}
%   \setlength{\tabcolsep}{3pt}
%   \caption{Ablation studies on different modules proposed in our method. The BASE model is a transformer, MOE, MMA, MMR denote f-MoE Adapter, multimodal Alignment and multimodal Reconstruction Modules, respectively. }
%   \label{table:2}
%   \centering
%   \begin{tabular}{ccccccc}
%     \toprule
%     Trial & Method & accuracy & AUC & sensitivity & specificity & F1  \\
%     \midrule
%     1 & \texttt{BASE} & 0.600 & 0.577 & 0.312 & 0.842 & 0.432\\
%     2 & \texttt{BASE + MMA} & 0.600 & 0.572 & 0.250 & 0.894 & 0.363\\
%     3 & \texttt{BASE + MMA + MMR} & 0.628 & 0.603 & 0.312 & 0.894 & 0.434 \\
%     4 & \texttt{BASE + MOE} & 0.771 & 0.764 & 0.687 & 0.842 & 0.733 \\
%     5 & \texttt{BASE + MMR} & 0.657 & 0.659 & 0.687 & 0.631 & 0.647 \\ %0.657 0.659 0.631 0.687 0.666
%     6 & \texttt{BASE + MOE + MMA} & 0.800 & 0.791 & 0.687 & 0.894 & 0.758 \\
%     7 & \texttt{BASE + MOE + MMR} & 0.800 & 0.810 & 0.937 & 0.684 & 0.810 \\
%     8 & \texttt{BASE + MOE + MMA + MMR} & \textbf{0.828} & \textbf{0.827} & 0.842 & 0.812 & \textbf{0.812}\\
%     \bottomrule
%   \end{tabular}
% \end{table}

\textbf{Analysis of the Semantic Embeddings.}
% Actually, the f-MoE Adapter, MMA and MMR modules are designed as a whole to obtain fine-grained representations for better classification performance. To better understand the representations learned by our method, we use the T-SNE map to visualize the semantic embeddings. In Fig\ref{fig2}, the red points are subjects diagnosed as MCI while the blue points are HCs. We can see the representations learned by our method have a clear clustering structure where blue points and red points are separated from each other, which makes it simple to discriminate the MCIs from the HCs.
% It is important that the f-MoE Adapter, MMA and MMR modules should be included in a complete manner. This will promote fine-grained representations for better classification performance. In Fig \ref{fig2}, we use t-distributed Stochastic Neighbor Embedding (\textit{t}-SNE) maps to visualize and for better understanding of the learned semantic embeddings by our method. It is clear that the representations learned by our method lead to much separated groups, indicating a better discriminative ability.
It is important that the f-MoE Adapter, MMA and MMR modules should be included in a complete manner. This will promote fine-grained representations for better classification performance. In Fig. \ref{fig2}, we use \textit{t}-SNE maps to visualize for understanding the learned semantic embeddings. It is clear that the representations learned by our method lead to much separated groups, indicating a better discriminative ability.

% \begin{table}
%   \setlength{\tabcolsep}{5pt}
%   \caption{Further discussion about the multimodal Alignment Module. }
%   \label{table:2}
%   \centering
%   \begin{tabular}{ccccccc}
%     \toprule
%     Trial & Method & ACC & AUC & SEN & SPE & F1  \\
%     \midrule
%     1 & \texttt{BASE} & 0.60 & 0.58 & 0.84 & 0.31 & 0.69\\

%     8 & \texttt{BASE + MOE + ALIGN + RECON} & \textbf{0.83} & \textbf{0.83} & 0.84 & 0.81 & \textbf{0.84}\\
%     \bottomrule
%   \end{tabular}
% \end{table}
\section{Conclusion}
This study presents an innovative research framework including sf-PET/MR, multimodal brain connectome construction and learning, clinically feasible multimodal fusion and diagnosis. The superior performance of MX-ARM is demonstrated by a precious, carefully curated sf-PET/MR dataset. The AUC of 0.741 also outperforms current SOTA performance in MCI detection, indicating that concurrent modeling brain metabolic, hemodynamic, and perfusion activity helps with more accurate early AD detection. %More importantly, MX-ARM has modal detachability and exchangeability, which enhances clinical flexibility, warranting its wide applications in the future.

\section*{Acknowledgment}
This work is partially supported by the STI2030-Major Project (No. 2022ZD0209000), the Shanghai Pilot Program for Basic Research - Chinese Academy of Science, Shanghai Branch (No. JCYJ-SHFY-2022-014), the National Natural Science Foundation of China (No. 62131015), the Shanghai Zhangjiang National Innovation Demonstration Zone Special Funds for Major Projects "Human Brain Research Imaging Equipment Development and Demonstration Application Platform"(No.ZJ2018-ZD-012), and the Shenzhen Science and Technology Program (No.KCXFZ20211020163408012). The data used in the present study are from the database constructed by Chinese Brain Molecular and Functional Mapping (CBMFM) project. 

\bibliographystyle{ieeetr}
\bibliography{ref}

\end{document}